\documentclass[prb,twocolumn,showpacs,preprintnumbers]{revtex4-2}%
\usepackage{graphicx}
\usepackage{epstopdf}
\usepackage{array,multirow,amsmath,amsfonts}
\usepackage{color}
\usepackage{amsmath}
\usepackage{amsfonts}
\usepackage{amssymb}%
\providecommand{\U}[1]{\protect\rule{.1in}{.1in}}
\begin{document}
\preprint{ }
\title{Competing spin-fluctuations in Sr$_{2}$RuO$_{4}$ and their tuning through epitaxial strain}
\author{Bongjae Kim$^{1}$}
\email{bongjae.kim@kunsan.ac.kr}
\author{Minjae Kim$^{2}$}
\author{Chang-Jong Kang$^{3}$}
\author{Jae-Ho Han$^{4}$}
\author{Kyoo Kim$^{5}$}

\affiliation{$^{1}$ Department of Physics, Kunsan National University, Gunsan 54150, Korea}
\affiliation{$^{2}$ Korea Institute for Advanced Study, Seoul 02455, Korea}
\affiliation{$^{3}$ Department of Physics, Chungnam National University, Daejeon, Daejeon 34134, Korea}
\affiliation{$^{4}$ Center for Theoretical Physics of Complex Systems, Institute for Basic Science (IBS), Daejeon 34126, Korea}
\affiliation{$^{5}$ Korea Atomic Energy Research Institute (KAERI), 111 Daedeok-daero, Daejeon 34057, Korea}
\date[Dated: ]{\today}

\begin{abstract}
 In this study, we report the magnetic energy landscape of Sr$_{2}$RuO$_{4}$ employing generalized Bloch approach within density functional theory. We identify the two dominant magnetic instabilities, ferromagnetic and spin-density-wave, together with \emph{other} predominant instabilities. We show that epitaxial strain can change the overall magnetic tendency of the system, and tune the relative weight of the various magnetic instabilities in the system. Especially, the balance between spin-density-wave and ferromagnetic instabilities can be controlled by the strain, and, eventually can lead to the new magnetic phases as well as superconducting phases with possibly altered pairing channels. Our findings are compared with previous theoretical models and experimental reports for the various magnetic features of the system, and offers first-principles explanation to them.
\end{abstract}
\keywords{engineering oxides}\maketitle

%%%%%%%%%%%%%%%%%%%%%%%%%%%%%%%%%%%%%%%%%%%%%%%%%%%%%%%%%%

%%%%%%%%%%%%%%%%%%%%%%%%%%%%%%%%%%%%%%%%%%%%%%%%%%%%%%%%%

\section{Introduction}
 Since the first report, Sr$_{2}$RuO$_{4}$ has long been an system of interests due to its unconventional superconducting properties~\cite{Maeno1994}. After intensive discussions on the superconductivity of the system, especially in the expectation of triplet pairing~\cite{Mackenzie2003,Maeno2012,Kallin2012,Liu2015,Mackenzie2017}, recent re-examination of the nuclear magnetic resonance experiments strongly suffocated the possibility of the chiral triplet pairing scenario~\cite{Pustogow2019}. This, however, renewed the interests in the system, by permitting other order parameters, and new candidates float up from both theoretical and experimental studies~\cite{Agterberg2020,Ghosh2020,Benhabib2020,Suh2020,Kivelson2020}.

 Unlike other highly-studied transition metal-based superconductors, such as cuprates and Fe-pnictides, Sr$_{2}$RuO$_{4}$ does not order magnetically and remains as a paramagnetic metal in its normal state down to very low temperature. But the system is known to be very close to the magnetic phase, and small perturbation, such as doping, easily turns the system into magnetic~\cite{Braden2002_1,Carlo2012,Ortmann2013,Kunkemoller2014,Zhu2017}. The leading magnetic instabilities of the system are known to be ferromagnetic (FM) and spin-density-wave (SDW) with $\mathbf{q} \sim (0.3,0.3,0)\frac{2\pi}{a}$ ($\mathbf{q}_{SDW}$)~\cite{Imai1998,Sidis1999,Braden2002_2,Iida2011}. The competition of the two magnetic instability is important as each can lead to different types of pairing symmetry~\cite{Mazin1997,Mazin1999,Eremin2004,Raghu2010,Tsuchiizu2015}. Recent polarized inelastic neutron scattering study has revealed the dominance of the SDW contribution in the spin-fluctuation spectrum over FM one~\cite{Steffens2019}, hence, putting even more weight on the singlet scenario. But if one can control the relative balance of the two competing magnetic instabilities, from the simple spin-fluctuation mediates pairing picture, we can expect the eventual tuning of the different types of the superconductivity pairing channels. Here, we pursue this direction through the strain engineering.

 In fact, the magnetism of Sr$_{2}$RuO$_{4}$ is not simple. In addition to the two leading instabilities, FM and SDW, many magnetic responses with different $\mathbf{q}$s are noticed~\cite{Braden2002_2,Eremin2002,Wang2013,Cobo2016,Liu2017}. Especially, the Fermi surface nesting instability at $\mathbf{q} \sim (1/2,1/4,0)\frac{2\pi}{a}$ is expected to promote the odd-parity pairing in uniaxially strained case~\cite{Roemer2020}. From the density functional theory (DFT) approach, susceptibility calculations readily reproduced the reported SDW instabilities, and their combination with many-body technique have offered further insights~\cite{Mazin1999,Boehnke2018,Strand2019}. But the energetics of the various magnetic phases of the system, which unambiguously discloses the instabilities, has not been presented from the DFT approaches. The magnetic energy scale of the Sr$_{2}$RuO$_{4}$ is known to be well-described by the DFT calculations despite the overestimating tendency towards magnetism~\cite{BKim2017}. Hence, the investigation of the stability of the magnetism in the broader ranges of the Brillouin zone will definitely lead to the better understanding of the magnetism and, consequently, of the superconductivity in Sr$_{2}$RuO$_{4}$.

 In this paper, we investigate the magnetic energy landscape of the Sr$_{2}$RuO$_{4}$ along the key $\mathbf{k}$-paths with various $\mathbf{q}$-values. We address the impact of FM and $\mathbf{q}_{SDW}$ fluctuations on the electronic structures of the system, and look for the prospect of controlling the leading magnetic instabilities by employing the epitaxial strain. Also, we search for the possible involvement of other types of the magnetic fluctuations.

\section{Calculation details}
 All calculations are performed employing the Vienna \emph{ab initio} simulation package (VASP)~\cite{Kresse1993,Kresse1996}. Generalized gradient approximation (GGA) by Perdew-Burke-Ernzerhof is utilized for the exchange-correlation functional~\cite{Perdew1996}. The energy cut for the plane waves of 600 eV was used with a Monkhorst-Pack $k$-mesh of 8 $\times$ 8 $\times$ 4. For the spin spiral calculations with various $\mathbf{q}$-values, we employed generalized Bloch condition as implemented in VASP~\cite{Sandratskii1991}. We have performed the full atomic relaxation for non-strained case (0\% strain). For the epitaxial strain simulation, we have fixed the in-plane lattice parameters based on the non-strained one and fully relaxed the other degree of freedoms.

%%%%%%%%%%%%%%%%%%%%
\begin{table}[b]
\caption{Comparison of supercell calculation with spin-spiral calculation employing generalized Bloch condition. For supercell calculation, magnetic structures were explicitly imposed with $\mathbf{q} = (1/3,1/3,0)\frac{2\pi}{a}$ ($\mathbf{q}_{1/3}$). For spin-spiral calculation, the most close $\mathbf{q}$ value with $\mathbf{q} = (0.33,0.33,0)\frac{2\pi}{a}$ is ($\mathbf{q}_{0.33}$) compared. The values in parenthesis are ones with the spin-orbit coupling calculations. $\mathbf{q}_{1/3}$-udd and $\mathbf{q}_{1/3}$-spiral denote the collinear and spiral configuration each with the same $\mathbf{q}_{1/3}$~\cite{BKim2017}. Unit of the energy difference is meV/Ru.}
\label{table1}
\begin{ruledtabular}
\begin{tabular}{c | c c c | c c}
        &  \multicolumn{3}{c}{supercell}  & \multicolumn{2}{c}{spin-spiral} \\
 (meV)  & FM &  $\mathbf{q}_{1/3}$-udd & $\mathbf{q}_{1/3}$-spiral & FM & $\mathbf{q}_{0.33}$ \\
\hline
Energy & 0.0 & -20.1 (-20.2) & -18.7 (-17.4)  & 0.0  & -28.5 \\
\end{tabular}
\end{ruledtabular}
\end{table}
%%%%%%%%%%%%%%%%%%%%%%%%%%%

 To check the validity of our approaches, we first compared the energetics of spin-spiral calculation with one from the explicit supercell calculation. The anisotropic terms are much smaller than the isotropic ones, which is already identified from the previous calculations~\cite{BKim2017}. Indeed, from the supercell calculations, the energy difference between collinear (udd:up-down-down) and spiral spin configurations with the same $\mathbf{q} = (1/3,1/3,0)\frac{2\pi}{a}$  ($\mathbf{q}_{1/3}$) are found to be 1.4 meV/Ru, which is much smaller than the energy difference between FM and $\mathbf{q}_{1/3}$ (around 20 meV/Ru) (Table.~\ref{table1}). This shows that the long-range magnetic periodicity, as represented by the propagation vector $\mathbf{q}$, determines the overall magnetic energy landscape rather than the local anisotropic details. Then, we compared our spin-spiral calculation employing the generalized Bloch condition for approximate $\mathbf{q}_{1/3}$, $\mathbf{q} = (0.33,0.33,0)\frac{2\pi}{a}$ ($\mathbf{q}_{0.33}$). The result shows that while there are small overestimation, the energy difference between FM and $\mathbf{q}_{0.33}$ phases is in similar scale (28.5 meV/Ru) as for the explicit supercell case. The overall magnetic energy scale is not changed much with the inclusion of the spin-orbit coupling (Table~\ref{table1}). With this, we can safely employ our spin-spiral calculation using generalized Bloch condition without spin-orbit coupling for the description of magnetic energetics. As noted before, the spin-spiral calculation is very tricky and sometimes unstable for different volume cases~\cite{Marsman2002}, and we have found that for the compressive strain of -3\% and beyond, the energetics description is not reliable. Furthermore, at such ranges, the system loses the magnetism and falls into the nonmagnetic phase. This is expected from the previous calculation on monolayer ruthenates~\cite{BKim2020}.

\section{Results and discussions}

%%%%%%%%%%%%%%%%%%%%
\begin{figure}[t]
\centering
\includegraphics[width=.95\columnwidth]{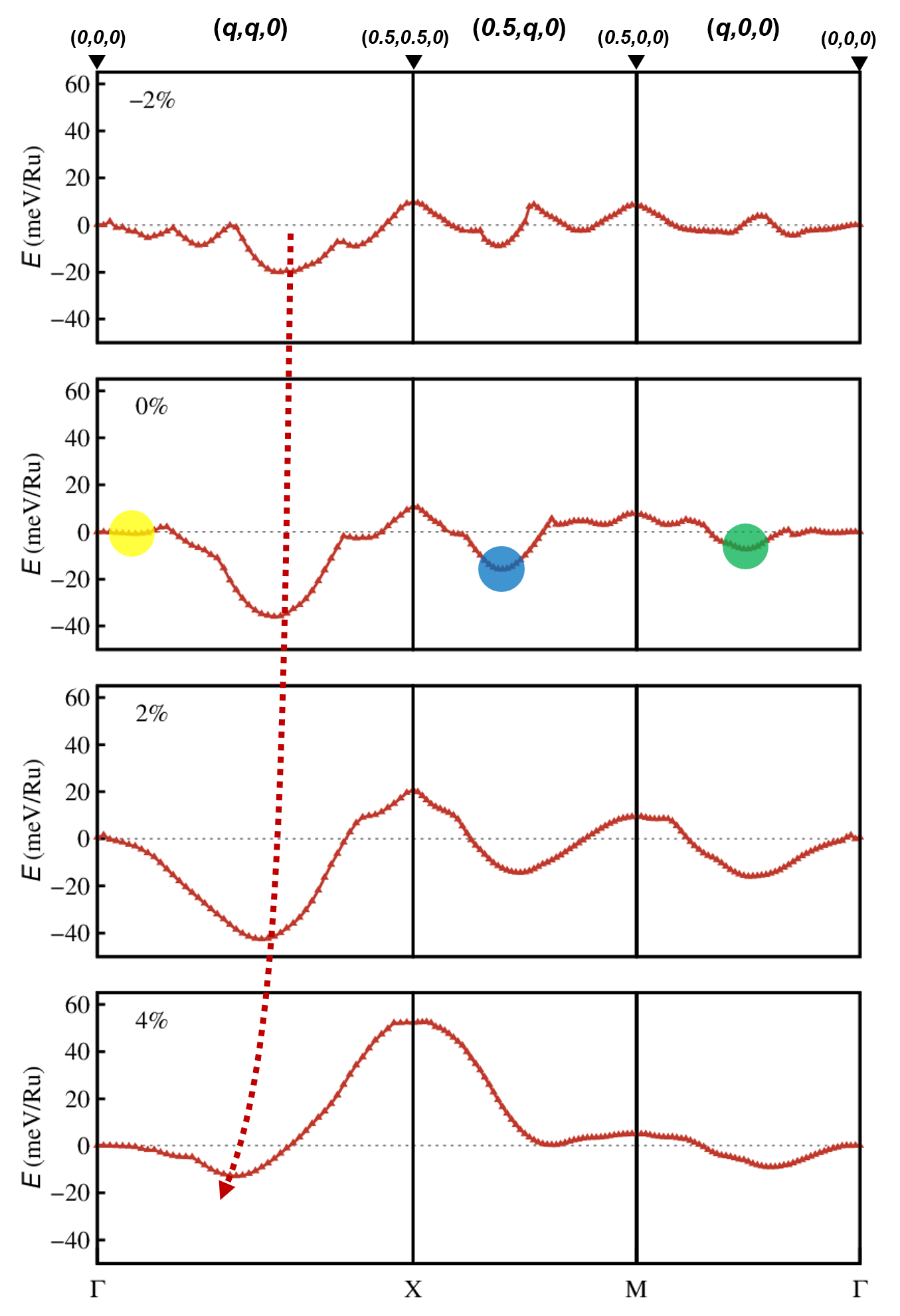}
\caption {
 Energetics of Sr$_{2}$RuO$_{4}$ upon various spin-spiral $\mathbf{q}$ wave-vectors along $\Gamma-X-M-\Gamma$. 0$\%$ strain corresponds to unstrained case, and plus and minus $\%$ to tensile and compressive strain, respectively. The colored circles denote the local minima for unstrained one (see text).
}
\label{fig1}
\end{figure}
%%%%%%%%%%%%%%%%%%%%

 In Fig.~\ref{fig1}, we have plotted the energetics of spin-spiral calculations with wave-vector $\mathbf{q}$ along the symmetric two-dimensional path $\Gamma-X-M-\Gamma$ for various biaxial strain cases. Let us first discuss for the non-strained case (0$\%$). We found that the lowest energy is found at $\mathbf{q} = (0.28,0.28,0)$ in $\Gamma-X$, which corresponds to the experimentally observed nesting wave-vector ($\mathbf{q}_{SDW}$) at $\mathbf{q} \sim (0.3,0.3,0)$~\cite{Braden2002_2,Iida2011}. Our obtained $\mathbf{q}_{SDW}$ position is slightly shifted to the $\Gamma$ point when compared to experimental one, but gives reasonable description. As the strongest magnetic instability is located at $\mathbf{q}_{SDW}$, which corresponds to the Fermi surface nesting vector, the global minimum at this point can be expected. Broad flat region at around $\Gamma$ point indicates the FM instability, which is energetically much higher (36 meV/Ru) than the one for $\mathbf{q}_{SDW}$. The broad flat curve suggests the high density of spin-spiral states and related entropies at around the FM instability. Note that from the recent inelastic neutron scattering study, a broad signal related to the FM fluctuations is reported~\cite{Steffens2019}. This feature is in stark contrast to the sharp peak from the SDW wave vector, which dominates the spin-fluctuation spectrum. In fact, we found that the FM instability is very fragile and tensile strain easily breaks the metastability at around $\Gamma$ as demonstrated in Fig.~\ref{fig1}. This indirectly explains why no FM magnetic order has been found in this compounds while various antiferromagnetic ones (including SDW) are found with external perturbations~\cite{Braden2002_1,Carlo2012,Ortmann2013,Kunkemoller2014,Zhu2017,Grinenko2021}.

 Other than the two leading magnetic fluctuations, SDW and FM, we recognize other instabilities from the energetics curve. First, near the $\Gamma$, the local minimum is found at $\mathbf{q} = (0.06,0.06,0)$ within the flat region (highlighted with yellow circle). Second, another meta-stable energy position is identified at $\mathbf{q} \sim (1/2,1/4,0)$ along $X-M$ (blue circle in Fig.~\ref{fig1}). From the previous three-band RPA susceptibility calculations, Cobo \emph{et al.} found magnetic fluctuations at $\mathbf{q} = (1/12,1/12,0)$ and $\mathbf{q} = (1/2,1/4,0)$ - the former corresponds to isotropic fluctuation and the latter to anisotropic intraband fluctuations~\cite{Cobo2016,Roemer2020}. Each can be compared reasonably to the local minima in our calculations. The $\mathbf{q} = (1/2,1/4,0)$ instability is not as dominant as $\mathbf{q}_{SDW}$ one, but is the next dominant feature. This instability is known to be highly enhanced upon uniaxial strain, upon the crossing of the $\gamma$-Fermi sheet through van Hove singularity (vHs)~\cite{Roemer2020}. While not discussed much, this magnetic reponse is also reported from the neutron scattering experiment~\cite{Iida2011}. A dip along $M-\Gamma$ path, marked in a green circle in Fig.~\ref{fig1}, is related to the nesting grid~\cite{Cobo2016,Strand2019,Iida2011}. Eremin \emph{et al.}, also reported the FM-related susceptibility signal at $\mathbf{q} = (0.1,0,0)$, which is within the broad region at around $\Gamma$ along $M$-$\Gamma$ path~\cite{Eremin2002}. The existence of various $\mathbf{q}$-wave vectors from previous susceptibility as well as our thorough energetics calculations directly shows the complex magnetic structure and instabilities involved in Sr$_{2}$RuO$_{4}$.

%%%%%%%%%%%%%%%%%%%%
\begin{figure}[t]
\centering
\includegraphics[width=.95\columnwidth]{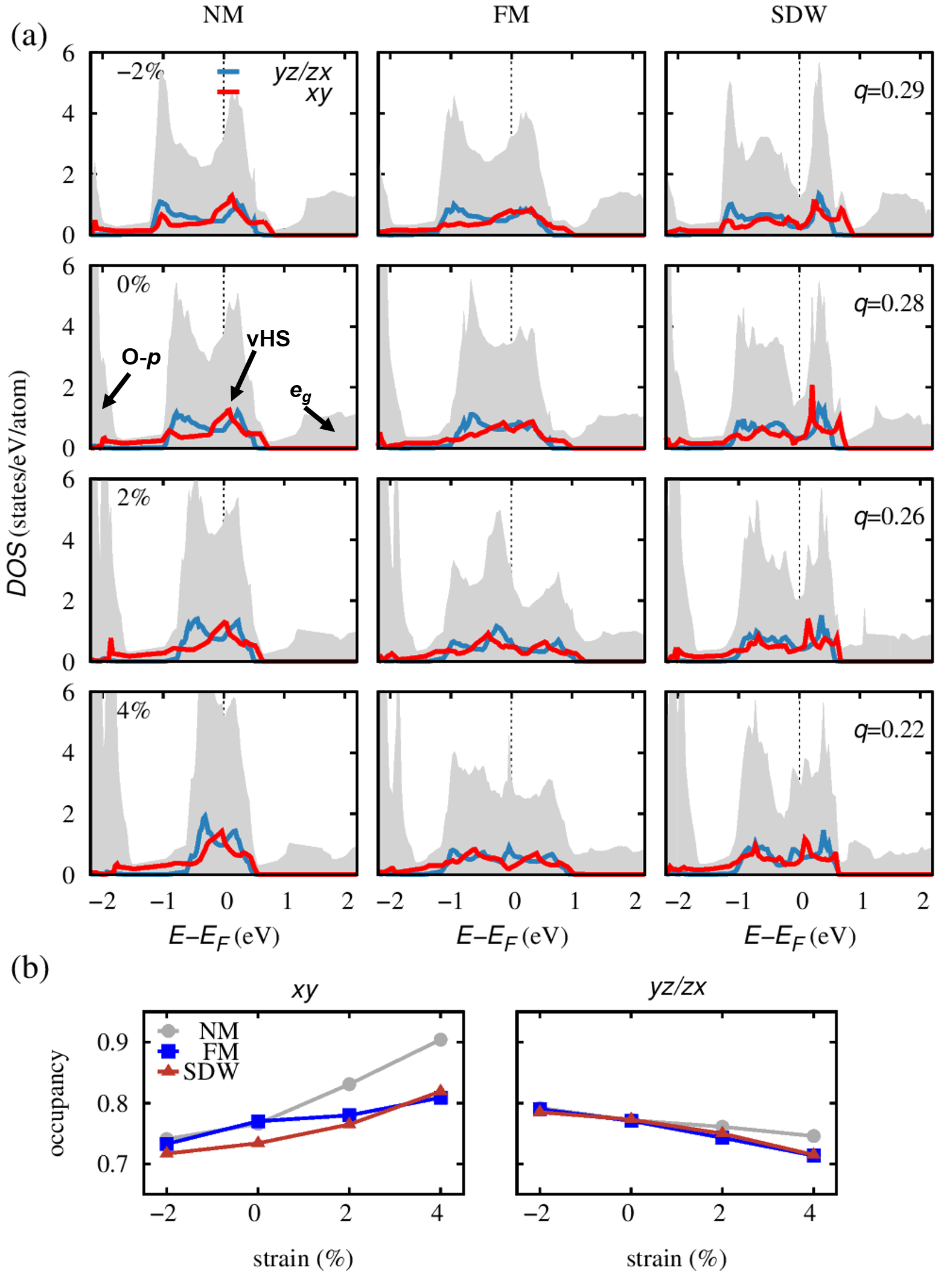}
\caption {
 (a) Total and orbital-resolved partial DOS of Sr$_{2}$RuO$_{4}$ for NM, FM, and SDW cases. For SDW, we have plotted the case of $\mathbf{q}_{SDW}$ with the minimum energy position for each strain cases. Gray color denotes the total DOS, red and blue/green colors are for $xy$- and $yz/zx$-orbital resolved partial DOS of Ru-$d$.
 (b) The orbital-resolved occupancy for Ru-$xy$ and $yz/zx$. The occupancy is obtained by integrating the partial DOS of each orbitals from -3.0 eV to E$_F$.
}
\label{fig2}
\end{figure}
%%%%%%%%%%%%%%%%%%%%

 The epitaxial strain strongly affects the overall shape of the energetics curve. Recent intensive studies employing the uniaxial strain, which breaks the $C_4$ symmetry, have provided unique avenue to understand the superconducting order parameter of the system~\cite{Hicks2014,Steppke2017,Grinenko2021}. The biaxial strain, usually exercised by epitaxially growing the Sr$_{2}$RuO$_{4}$ on top of typical substrates such as SrTiO$_3$ provides stable route to control the electronic and magnetic properties compared to the difficult uniaxial approaches~\cite{Barber2019}. The tensile strain moves the $\mathbf{q}_{SDW}$-vector towards $\Gamma$ point (see dotted arrow in Fig.~\ref{fig1}), which, as we will see below, directly indicates the changes in the Fermi surface nesting feature. Simultaneously, the strain tunes the relative stability of FM and SDW phase. Starting from -2\%, as the system is strained, SDW is more stabilized up to 2\%, where the energy difference with FM one is as high as 43 meV/Ru. Then for the extreme case of 4\% strain, the energy difference between FM and SDW is greatly reduced to 13 meV/Ru. This suggests that through epitaxial engineering, the relative dominance of SDW over FM can be changed, and, in turn, can change the spin-fluctuation pairing channels. Considering the doping stabilizes the magnetic ground states, we can also expect the epitaxial strain to play the similar role. Note that the subdominant $\mathbf{q} = (1/2,1/4,0)$ instability corresponding to the one marked in the blue circle in Fig.~\ref{fig1}, which is anisotropic, cannot be stabilized upon biaxial strain. If one breaks the $C_4$ symmetry through uniaxial strain, the magnetic interaction between the next nearest neighbor Ru bifurcates, which can promote the anisotropic spin fluctuations~\cite{BKim2017,Roemer2020}. Still, the dominant one is the SDW-fluctuations~\cite{BKim2022}.

%%%%%%%%%%%%%%%%%%%%
\begin{figure*}[t]
\begin{center}
\includegraphics[width=\textwidth]{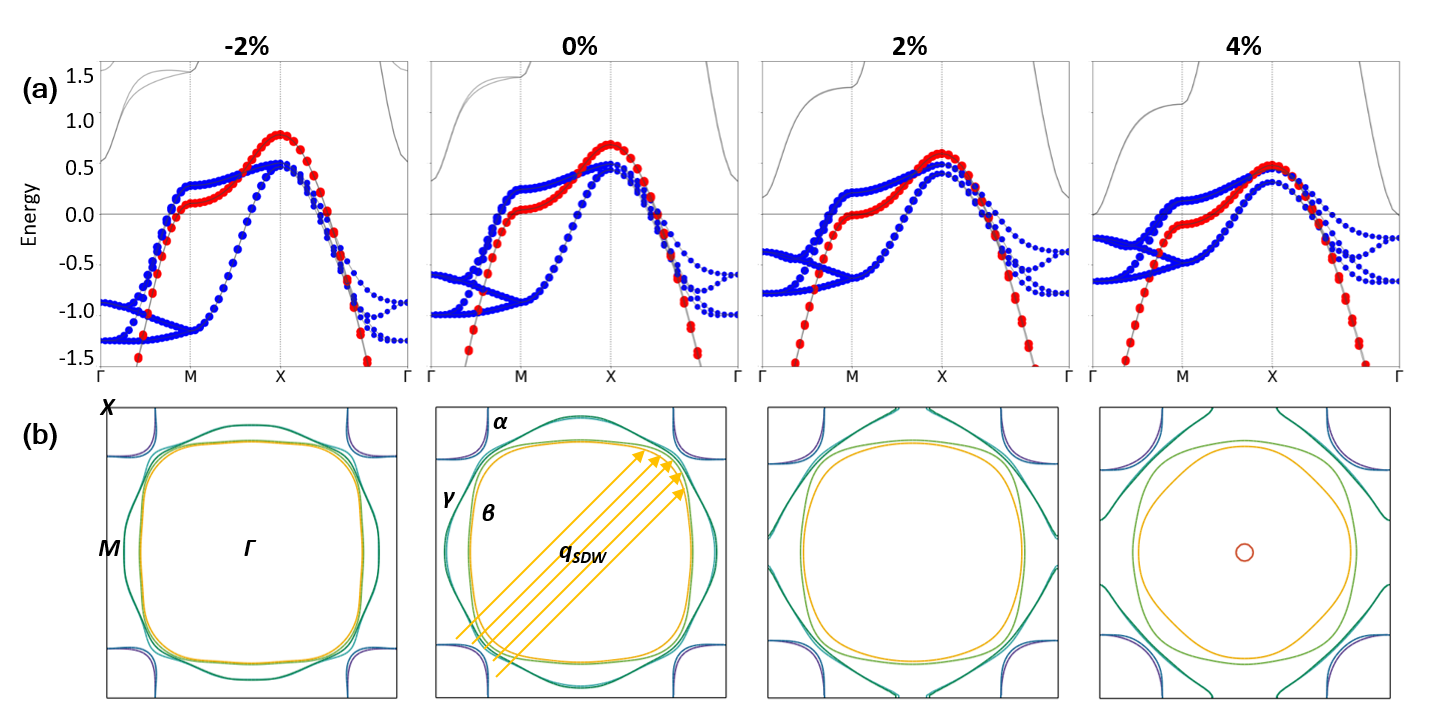}
\end{center}
\caption{(Color online) (a) The band structure and (b) Fermi surface plot for various strain cases. For band structure, $xy$- and $yz/zx$-projected bands are denoted with red and blue colors. The $\alpha$ and $\beta$ Fermi surfaces are quasi-one dimensional with $yz/zx$ character and $\gamma$ one is two-dimensional with $xy$ character. The arrows in the Fermi surface indicates $\mathbf{q}_{SDW}$ nesting vector which activates the quasi-1D Fermi surfaces
}
\label{fig3}
\end{figure*}
%%%%%%%%%%%%%%%%%%%%

 The magnetic energy landscape is smoothed upon tensile strain. The local minimum at $\Gamma$ quickly disappears and for extreme case of 4\% strain, the relevant energy scales among local minima become very small, suggesting the dominance of SDW is greatly reduced. In fact, the overall magnetic tendency is enhanced upon the tensile strain. We found the local magnetic moment increases upon the tensile strain, which is expected as the tensile strain enlarges the bondlength between Ru sites, hence, the system becomes more localized. Compared to the complex energetic landscape for compressive case, involved magnetism for 4\% strain case is incomplex. At this regime, the instabilities are very close in energy, and the dominance of nesting-induced SDW fluctuation is almost collapsed. As we will discuss below, strong tensile strain changes the morphology of the Fermi surface and weakens the nesting feature. For compressive strain, the overall energy scales are also reduced. But this is not due to the enhanced competition among differernt magnetism but from the suppression of the overall magnetism due to the increased itinerancy. The local magnetic moment of Ru at $\mathbf{q}_{SDW}$ is 0.78$\mu_B$ for -2\% strain case, which is much smaller than the corresponding value 1.25$\mu_B$ for 4\% strain. We found that the compressive strain beyond -2\% kills the magnetism of the system even with the overestimating tendency of the GGA functionals towards magnetism~\cite{BKim2017}.

 To investigate the role of magnetic fluctuations on the electronic structures, in Fig.~\ref{fig2}(a), we have plotted density of states (DOS) of nonmagnetic (NM), FM, and SDW phases for each strain cases. Well-known electronic structures of Sr$_{2}$RuO$_{4}$ can be seen for unstrained case (See NM 0\% case of Fig.~\ref{fig2}(a)): The vHs peak, which is mainly from Ru-$xy$ orbital, is located slightly above the Fermi level (see arrow) and the broad two-peak structures from Ru-$zx/yz$ orbitals are well-reproduced. As the electronic structures of layered ruthenates are very close to the Stoner instability, magnetism is expected to relieve the high DOS at the Fermi level, and in this case, also the vHs peak~\cite{Mazin1997}. Here, the two leading magnetism, FM and SDW, act very differently. In the case of FM, the vHs from the $xy$-orbital is relieved and the peaked structure is not remained. Hence the $xy$-orbital contribution at the Fermi level is reduced. But FM does not significantly alter the structure of the $yz/zx$-orbitals, indicating the FM fluctuation is strongly tied to the $xy$-orbital. While the vHs peak is dissipated, the DOS \emph{at} the Fermi level is not changed much, so is the Stoner instability. For the SDW case, however, one can see the strong suppression of the DOS at the Fermi energy with pseudogap-like feature. Differently from the FM case, this feature is contributed from all three $t_{2g}$ orbitals. Our findings are in good accordance with previous susceptibility calculation, where $\Gamma$-point fluctuation is mostly from the $xy$-orbital but the SDW one is contributed from all three orbitals~\cite{Strand2019}. Furthermore, while FM suppresses the vHs from the $xy$-orbitals, SDW shifts the position of vHs to slightly higher energy without relieving the peak itself. Hence, vHs feature is preserved even with the SDW fluctuation. Recent study have found the emergence of the SDW order well beyond the vHs crossing strain~\cite{Grinenko2021}. Our calculations suggest the robustness of the van Hove peak upon the static magnetic order, which can be an interesting feature and requires further investigations.

 Despite the marked differences in partial DOS of FM and SDW, the overall orbital-resolved occupancies do not show much distinction. In Fig~\ref{fig2}(b), we display the occupation of each orbital by simply integrating the partial DOS from -3 eV below to the Fermi energy. While the value itself has an ambiguity due to the hybridization with O-$p$ orbitals, we can clearly observe the expected increasing and decreasing tendency in the occupation of $xy$ and $yz/zx$ orbitals, respectively, as the tensile strain is applied. As shown in Fig.~\ref{fig2}(b), due to the tetragonal crystal field, the occupancy of the $xy$ orbitals increase upon tensile strain, and $yz/zx$ orbitals show the opposite tendency. Here, we note the different types of magnetic orders do not have much impact on the occupancy. This may suggest the magnetism itself does not entangle much with the orbital-dependent electronic behaviors such as orbital-selective Mott phase~\cite{Koga2004}.

 The general electronic structures of FM and SDW phase are not changed much upon strain. Interestingly, the tensile strain moves the vHs peak towards the Fermi level, which is similar to previous NM calculations. As previous studies have shown, the superconducting critical temperature can increase upon the epitaxial strain both from the singlet and triplet picture~\cite{Burganov2016,Hsu2016,Liu2018}. Hence, we believe the direct consideration of magnetic fluctuations in the Hubbard-Kanamori type approaches can offer more insight in this system~\cite{BKim2017}.

 As the tensile strain is applied, we first see the overall bandwidth of the Ru-$t_{2g}$ orbitals are progressively decreasing. Especially the $xz/yz$-orbitals strongly respond to the strain compared to $xy$-orbital. Also, upon tensile strain, we can see that the contribution of Ru-$e_g$ and O-$p$ is enhanced. From partial DOS in Fig.~\ref{fig2}, Ru-$e_g$ (O-$p$) orbitals move down (up) towards the Fermi energy. The low-energy physics of Sr$_{2}$RuO$_{4}$ is commonly based on the three-band ($t_{2g}$ orbitals) picture, which may require modifications for the tensile strain case due to the involvement of $e_g$ orbitals at around the Fermi energy.

 In Fig.~\ref{fig3} (a) and (b), we plotted the NM band structures and Fermi surfaces for each strain case~\cite{Ganose2021}. Ru-$e_g$ bands progressively shift down and touches the Fermi level at $\Gamma$-point for 4\% strain case. We clearly see the bandwidth of $xz/yz$-orbitals are strongly narrowed upon strain in contrast to the mild change of the $xy$-orbitals (see Fig.~\ref{fig3}(a)), as already noted from the DOS in Fig.~\ref{fig2}(a). From the Fermi surface, 1-dimensional $\alpha$ pocket enlarges upon the tensile strain, which changes the overall size of the nesting-vector and explains the $\mathbf{q}_{SDW}$ movement from $X$ to $\Gamma$ in Fig.~\ref{fig1}. Noteworthy is that the abrupt change of $\mathbf{q}_{SDW}$ vector in between 2 and 4\% tensile strain is further assisted by the $\gamma$ band. The Lifshitz transition of the $\gamma$ band is accompanied by the abrupt change in the velocity of curvature in the bands (See red colored band in Fig~\ref{fig3}(a)) and the diamond-shape Fermi sheet can contribute to the electronic susceptibility as in the 1-dimensional pockets. We can also notice that the morphology of the $\alpha$ and $\beta$ Fermi surfaces is severely distorted. Especially, the $\alpha$ pocket is progressively changed from squared to rounded shape. This weakens the nesting effects, and eventually destabilize SDW as we see from the energetics landscape in Fig.~\ref{fig1} and the evolution of the Fermi surface as a function of the strain displayed in Fig.~\ref{fig3}(b). The tensile strain moves the $\gamma$ pocket edge, which crosses the $M$-point at 2\% strain. The shape of the $\gamma$ sheet changes from circular to rhombic shape, which indicates the enhanced $\sigma$-bonding over $\pi$-bonding. For 4\% strain, we can see the contribution from the $e_g$ orbitals, at $\Gamma$. At this limit, the low-energy physics of the system cannot be accounted with $t_{2g}$-only three-band model.

\section{Conclusions}

 In conclusion, employing spin-spiral DFT energy calculations, we have identified the magnetic energy landscape of Sr$_{2}$RuO$_{4}$. We investigated the evolution of the magnetic instabilities by biaxially applying the epitaxial strain to the system. We found the fragile character of FM fluctuations and the robust character of SDW one. The latter remains as the most stable one for all strain ranges. However, their relative instability can be tuned upon external strains, which can be important for the identification of dominant pairing fluctuations. We further report the various other magnetic instability, which were identified in previous theoretical models as well as experiments. While the current studies on the symmetry-breaking uniaxial strain can offer key insight to understand the magnetism and superconductivity of the system, we envisages that the biaxial strain can be a functional way to tune the magnetism, and, eventually the superconducting pairing channel of the system. We expect energetics studies with explicit inclusion of the dynamic correlation, such as DFT plus dynamical mean-field theory, can further offer valuable insights.

\section{Acknowledgements}
 The authors like to acknowledge the support from Advanced Study Group program from PCS-IBS. B.K. acknowledges support by NRF grant No. 2021R1C1C1007017 and KISTI supercomputing Center (Project No. KSC-2021-CRE-0605). M.K was supported by KIAS individual Grants No. CG083501. C.J.K. was supported by the NRF grant No. 2022R1C1C1008200 and the National Superconducting Center with supercomputing resources including technical support KSC-2021-CRE-0580. J.H.H. acknowledge financial support from the Institute for Basic Science in the Republic of Korea through the project IBS-R024-D1. K.K. acknowledges support from KAERI Internal R\&D program (No. 524460-22).

\bibliographystyle{apsrev4-2}
\bibliography{bibfile}

%\newpage
%\input{sm_2020}

\end{document}